\documentclass[epj,final]{svjour}
\usepackage{graphicx}
\def\be{\begin{eqnarray}}
\def\ee{\end{eqnarray}}
\def\bc{\begin{center}}
\def\ec{\end{center}}

\def\rmd{{\rm d}}
\def\om{\omega}

\newcommand{\lsim}{\stackrel{\scriptstyle <}{\phantom{}_{\sim}}}

\begin{document}
\title{Complete Strangeness Measurements in Heavy-Ion Collisions}
\author{Boris Tom\'a\v{s}ik\inst{1} and  Evgeni E.~Kolomeitsev\inst{1}}
\institute{Univerzita Mateja Bela, FPV, Tajovsk\'eho 40,
97401 Bansk\'a Bystrica, Slovakia
 \and Czech Technical University in Prague, FNSPE, B\v{r}ehov\'a 7,
11519 Prague 1, Czech Republic}
\date{}
\abstract{
We discuss strangeness production in heavy-ion collisions
within and around the energy range of the planned NICA facility.
We describe the minimal statistical model, in which the total strangeness yield is fixed by the observed or calculated $K^+$ multiplicity. We show how the exact strangeness conservation can be taken into account on the event-by-event basis in such a model.
We argue that from strange particle yields one can reveal information about the collision dynamics and about possible modifications of particle properties in medium.
This can be best achieved if the complete strangeness measurement is performed, i.e.  kaons, anti-kaons, hyperons and multistrange hyperons are registered in the same experimental setup.
In particular the production of hadrons containing two and more strange quarks, like $\Xi$ and $\Omega$ baryons could be of interest.
}

\maketitle

\section{Strangeness as a messanger}

Strangeness production in heavy-ion collisions (HICs) is in the center of interest for a long time since the first strange particles were registered in Ne on NaF and Ar on KCl collisions at Bevalac~\cite{Harris-PRL47,Schnetzer-PRL49}. This can be explained by several reasons. First, strangeness can be seen as a tag on a hadron, indicating that this particular hadron most probably was produced in the course of a collision and does not stem from the original nuclei.
Second, strange quarks and anti-quarks prefer to reside in different hadronic species.
A strange quark prefers to be mainly in baryons, i.e.  in one of the several baryonic states
$\Lambda$, $\Sigma$, $\Xi$, $\Omega$ or in $\bar K$ mesons.
Oppositely, the anti-strange quark can only reside in mesonic states $K^+$, $K^0$.
Also for higher mass resonances there are many more hadronic states available where strange quarks can be accommodated than there are for antiquarks.
Such an asymmetry restores if one takes into account anti-baryons, but at NICA energies the anti-baryon admixture is small.
As the result strange and anti-strange subsystems interact differently with the non-strange baryon-rich environment. The anti-strange subsystem couples weaker than the strange one. As argued in~\cite{Ko83,KVK96,Greiner91} this could leads to strangeness/anti-strangeness separation in HIC at energies from SIS to middle RHIC energies.
Third, because strangeness is conserved in strong interactions, strange particles are created in elementary collisions only being accompanied by anti-strange particles. Therefore, the strangeness production threshold is high. This makes yields of strange particles in HIC
sensitive to possible in-medium effects.

\begin{figure}
\centering
\includegraphics[width=7.5cm]{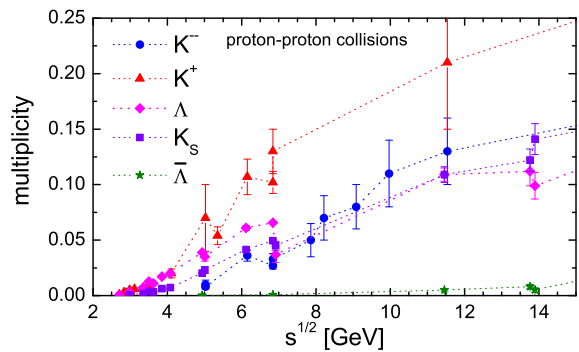}
\caption{Multiplicities of strange particles measured in proton-proton collisions as functions of the center-of-mass energy~\cite{GR96}.}
\label{fig:Sinpp}
\end{figure}

The study of strangeness in HIC is complicated however by several factors. First of all, cross sections of reactions with the strangeness production are poorly known, especially close to the reaction threshold. Experimentally accessible are only reactions with the $pp$, $np$ and $\pi p(n)$ entrance channel, like those currently studied, e.g., by HADES collaboration at GSI~\cite{HADES} and by ANKA, HIRES, and TOF collaborations at COSY~\cite{COSY}.
Without reliable experimental information about reactions with the $nn$ initial state we have to rely on the isospin symmetry relations, making {\it ad hoc} assumptions about internal mechanisms of the reactions~\cite{Fischer}. Even for higher energies the experimental situation with the strangeness production in elementary reactions does not become much better. As an illustration we show in Fig.~\ref{fig:Sinpp} multiplicities of strange particles produced in $pp$ collisions at various $\sqrt{s}$. The multiplicities grow with an increase of $\sqrt{s}$ but error bars increase also.
The plot also does not exclude that in the region $4~{\rm GeV}< \sqrt{s}< 7~{\rm GeV}$ the $K^+$ production mechanism changes as the $\sqrt{s}$ dependence of the $K^+$ multiplicity changes rapidly therein.
The second factor complicating the strangeness study in HICs is that interactions of strange particles with non-strange ones and among themselves are not fully constrained. There are limited data in channels with total strangeness $S=\pm 1$, like a hyperon-nucleon and anti-kaon--nucleon scattering and no experimental data of hadron reactions with total strangeness
$S=-2$, $-3$.
Finally, the properties of strange particles and their interactions in dense and hot nuclear matter are far from being firmly established.

Against all the above mentioned complications a satisfactory description of the kaon and $\Lambda$ productions in HICs has been reached in various transport~\cite{Fuchs04,Hartnack12} and statistical~\cite{KVK96,Cleymans} approaches.


\section{Strange particles in medium}\label{sec:medium}

Medium effects are found to be important for the description of particle production in HIC at SIS energies~\cite{Revs,Voskre-HIC}.
In particular, strange particle yields are strongly influenced by them~\cite{Fuchs04,Hartnack12,KVK96,LLB97,BratCass}. For hyperons and nucleons the in-medium modification of the energy spectrum of particle $a$ is often effectively parameterized in terms of scalar $S_a$ and vector $V_a$ potentials $ E_a(p)=\sqrt{m_a^{*2}+p^2}+V_a$, where the scalar potential enters the spectrum through the effective mass $m_a^* =m_a+S_a $. The description in terms of the $S_a$ and $V_a$ potentials is typical for relativistic mean-field (RMF) models, cf.~\cite{Fuchs04,KV05}. The nucleon potentials determined in~\cite{KV05} are $S_N\simeq-190~{\rm MeV}\rho_B/\rho_0$ and $V_N\simeq + 130~{\rm MeV}\rho_B/\rho_0$, here
$\rho_B$ is the baryon density and $\rho_0=0.16/{\rm fm}^3$ is the nuclear saturation density. The same potentials could be also used for $\Delta$: $S_\Delta\simeq S_N$, $V_\Delta\simeq V_N$.

One usually relates the hyperon potentials to the nucleon ones, $V_H=\alpha_H V_N$ and
$S_H=\beta_H S_N$. The parameter $\alpha_H$ is chosen according to the number of non-strange quarks in the hyperon, $\alpha_\Lambda=\alpha_\Sigma=2\alpha_\Xi=2/3$.
The parameter $\beta_H$ is, then, chosen such that the optical potential of a hyperon in nuclear medium at saturation $U_H=S_H(\rho_0)-V_H(\rho_0)$ agrees with the empirical information from
the analysis of hypernuclei:  $U_\Lambda=-27$~MeV, $U_\Sigma=+24$~MeV, and $U_\Xi=-14$~MeV.

\begin{figure}
\bc
\includegraphics[width=7.2cm]{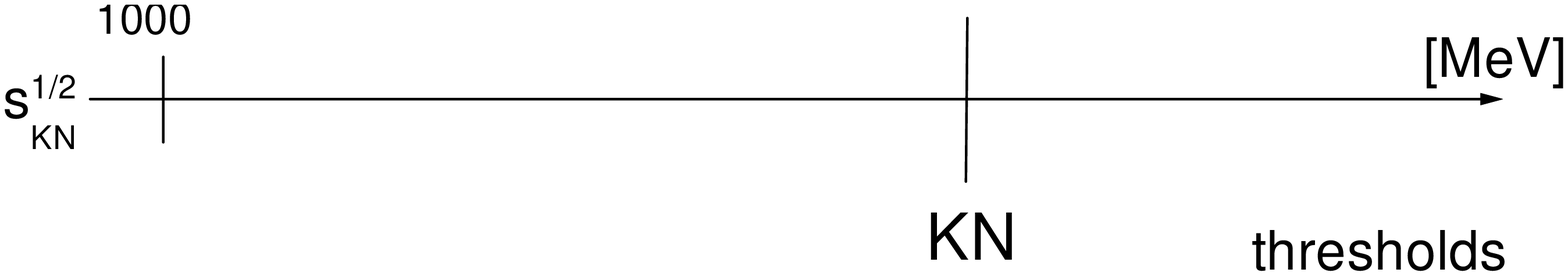} \\  
\includegraphics[width=7.2cm]{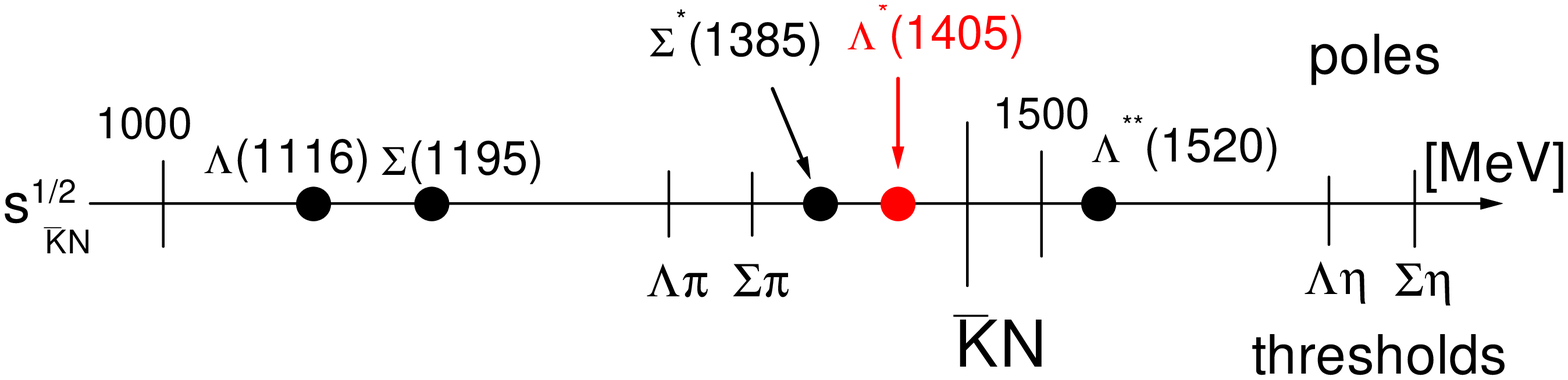}
\ec
\caption{World of kaon-nucleon (upper panel) and anti-kaon--nucleon (lower panel) scattering. Vertical dashes show thresholds and fat dots correspond to poles.}
\label{fig:meson-map}
\end{figure}
In the sector of strange mesons the situation is more complicated.
In Fig.~\ref{fig:meson-map} we present the maps of the worlds of $KN$ and $\bar{K}N$ scattering.
The $KN$ interaction is structureless without any resonances and elastic up to rather high energies.
The kaon self-energy
was evaluated in~\cite{LKK} in terms of the real parts of the s- and p-wave kaon--nucleon scattering amplitudes
obtained from the non-perturbative solution of the Bethe-Salpeter scattering equation~\cite{LK02}
with the kernel given by the chiral perturbation theory and parameterized as $\Pi_K^{\rm (NP)}(\om,\vec{q}\,) = \big(1.1\,m_K-\om+0.2\,\vec{q}\,^2/m_K\big)$ $ 46.8\,{\rm MeV}  \rho/\rho_0$\,, with $m_K$ being the kaon mass.
The in-medium kaon spectrum follows then from solution of the Dyson equation, $\om^2-\vec{q}\,^2-m_K^2- \Pi_K(\om,\vec{q}\,)=0$.
We note that the results obtained with the non-perturbative scattering amplitude differ from those obtained on a tree level from the leading-order chiral Lagrangian.

The world of antikaon--nucleon scattering is much richer: it is a coupled-channel system as the $\bar{K}N$ scattering is inelastic already on the threshold due to the coupling to $\pi\Lambda$ and $\pi\Sigma$ channels. There is a bunch of resonances, among which the s-wave $\Lambda(1405)$ resonance dominates the $\bar{K}N$ scattering in the isospin-0 channel close to the threshold. In the p-wave, the resonance $\Sigma(1385)$ determines the isospin-1 scattering amplitude.
The $\bar{K}N$ interaction is strongly attractive right below the threshold. This would lead to a shift of the $\bar{K}$ spectral density for given $\vec{q}$ in a nuclear medium towards lower energies. There, kaonic modes can couple strongly to the hyperon--nucleon-hole modes ($HN^{-1}$)~\cite{KVK96} built by ground-state $\Lambda$ and $\Sigma$ hyperons. The complicated  interplay of anti-kaons and hyperon-resonance propagation in medium asks for self-consistent calculations~\cite{LKorp02,Tolos06}. There are attempts to implement the  anti-kaon spectral functions in transport calculations~\cite{Cass-Tolos}.
As argued in~\cite{Ko83,KVK96} anti-kaons leave the fireball created in a HIC at the last stage of it evolution, when the baryon density is rather low $0.5\rho_0\lsim \rho_B\lsim\rho_0$. Therefore instead of the spectral function one frequently uses an effective scalar potential changing the $\bar{K}$ mass as $m_K^*=m_K+U_{\bar{K}}\rho_B/\rho_0$ with the parameter
$U_{\bar{K}}=-(70\mbox{--}150)$~MeV.


\section{Minimal statistical model for strange particles}

We consider a simplified picture of a HIC assuming that a thermalized (at least localy) nuclear system (a fireball) is formed. This system participates in a hydrodynamical expansion, which lasts until the moment of freeze-out characterized by the density $\rho_{B,\rm fo}$ and temperature $T_{\rm fo}$. Henceforth, in-medium particle thermal momentum distributions become distributions of free-streaming particles.  Because of the high production thresholds the strange particles are most efficiently produced in the first nucleon-nucleon collisions at the beam energies above the production threshold and at the early hot and dense stage of the fireball evolution.

At SIS-NICA-SPS energies the fireball is baryon-rich, and antistrange particles ($K^{+,0}$ mesons) have longer mean free paths than strange particles (antikaons and hyperons). Therefore kaons can easily move off the production point, and either leave the fireball immediately or first thermalize via elastic kaon-nucleon scattering and then leave at some intermediate stage. Because the population of strange particles is very small, even if a kaon stays in the fireball for a while, there is little chance that it meets an antikaon or hyperon and is absorbed by them. Thus in the course of collision the amount of strangeness in the fireball grows. The accumulated strangeness is redistributed among various hadronic species, like $K^-$, $\bar K^0$, $\Lambda$, $\Sigma$, $\Xi$, $\Omega$ and their resonances, which are released at the fireball breakup. Thus the kaon yield measured experimentally or calculated theoretically in a hadro-chemical model can be used to normalize the abundance of strange particles.

From the experimental point of view  {\it complete strangeness measurements}, i.e.\ measurement
of several strange species --- $K^+$, $K^-$, $\Lambda$ and $\Xi$ --- in one and the same experimental run, would allow to better understand strangeness production mechanisms, interrelations of strange particle observables, and a possible role of in-medium effects.

Relative abundances of strange particles in HIC can be calculated as in~\cite{KT05} within a hadronic kinetic model. Being interested only in the ratios of total yields it is enough to study the evolution of the spatially averaged densities of individual species only. If
we assume that the fireball is spatially uniform and we can characterize it by a time-dependent temperature $T(t)$, baryon density $\rho_B (t)$ and volume $V(t)$, then the time evolution of
the $K^+$ multiplicity, $M_{K^+}$,  or density, $\rho_{K^+}=M_{K^+}/V$  is determined by the equaiton
\begin{eqnarray}
\label{Kprod}
\frac{\rmd M_{K^+}}{\rmd \tau}=
\sum_{ij} V  \frac{\langle v_{ij}\sigma^+_{ij} \rangle }{1+\delta_{ij}} \rho_i\, \rho_j
- V
\mathcal{R}_{\rm loss}\,,
\end{eqnarray}
where $V$ is the volume of the system, $\tau$ is the proper time.
The first term on the right hand side is the kaon production rate with
the summation running over non-strange hadron species with densities $\rho_i$ and the
$\langle v_{ij}\sigma^+_{ij} \rangle$ standing for the reaction cross section averaged with the relative velocity of the colliding particles over the relativistic Boltzmann distributions.
The most important reaction channels with $\pi B$, $BB$ ($B=N,\Delta,Y$) and $MM$ ($M=\pi,\rho$) entrance channels are included.  The second term, $\mathcal{R}_{\rm loss}$ represents the rate of kaon annihilation processes. As demonstrated in~\cite{KT05} for AGS--SPS energies $\mathcal{R}_{\rm loss}$ can be neglected, so it can be done also for NICA energies. Equation (\ref{Kprod}) is integrated starting from some initial kaon density due to kaon production in incident nucleon collisions, which can be estimated from a compilation of kaon production data in nucleon--nucleon collisions~\cite{GR96}.

Since the reactions which do not create strange quarks but rather rearrange them in different hadrons are very quick, with a good approximation one can assume that all species containing strange quarks are in {\em relative chemical equilibrium}. The strangeness conservation imposes then the relation among the densities
$
\rho_S \equiv \sum_{i,\, S<0} |S_i|\, \rho_i = (1+\eta)(\rho_{K^+}  + \rho_{K^{*+}})
$
where the sum runs over all species containing strange quark. Here the parameter  $\eta=(A-Z)/Z$ quantifies the isospin symmetry breaking in HIC.

The outlined approach was applied in~\cite{KT05} for the description of the strange particle multiplicites in Au$+$Au collisions at AGS-SPS energies. There, the evolution of the energy density and the baryon density was parameterised with varying input parameters such as, e.g., the total fireball lifetime and the initial energy density. The energy density and number densities in the final state are chosen so that they correspond to the values extracted within the statistical model~\cite{Becattini04}. In~\cite{KT05} in-medium effects were included neither for hyperons nor for mesons, as they were deemed to be unimportant at the fireball temperatures and densities characteristic for this energy range.
\begin{figure}
\begin{center}
\includegraphics[width=7.4cm]{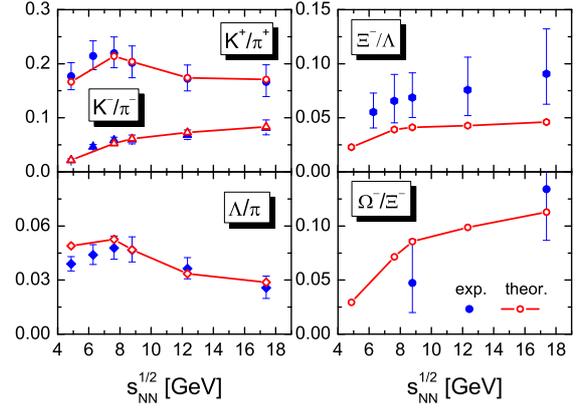}
\end{center}
\caption{Excitation functions of the strange particle ratios calculated in~\cite{KT05} for Au+Au collisions in comparison with the data~\cite{CERN-data,Lung}.}
\label{fig:SPS-ratios}
\end{figure}
The resulting particle ratios are depicted in Fig.~\ref{fig:SPS-ratios}.
Our hadronic kinetic model~\cite{KT05} is able to describe satisfactorily the excitation function of the ratios $K^+/\pi^+$, $K^-/\pi^-$, and $\Lambda/\pi$.
\begin{figure}
\begin{center}
\includegraphics[width=7.7cm]{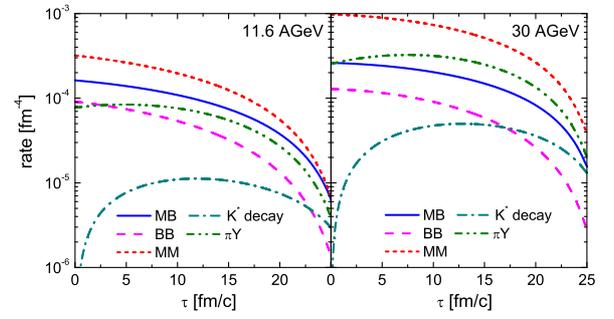}
\end{center}
\caption{$K^+$ production rates in meson-meson (MM), meson-baryon (MB), baryon-baryon (BB), pion-hyperon ($\pi Y$) collisions and in $K^*$ decays as functions
of time for scenarios which fit the data at: 11.6~$A$GeV,
30~$A$GeV}
\label{fig:Kpl-rates}
\end{figure}
Different contributions to the kaon production rate are shown in Fig.~\ref{fig:Kpl-rates} for Au+Au collisions at beam energies 11.6$A$GeV ($\sqrt{s}=4.86A$GeV) and 30$A$GeV ($\sqrt{s}=7.62A$GeV). Kaons are produced mainly in the MM reactions among which the
$\pi\pi$ channel dominates. The next in importance are the MB and $\pi Y$ channels. The contribution of the letter increases with an increase of the collision energy as the temperature of the fireball and consequently the hyperon concentrations grow.

In Fig.~\ref{fig:SPS-ratios} we give also the results for the ratios involving $\Xi$ and $\Omega$ baryons containing two and three strange quarks. The experimental data on $\Omega$ yields are still very scarce. The ratio $\Xi^-/\Lambda$ remains underestimated in the entire range of collision energies. This can indicate, perhaps, that the $\Xi$ baryons decouple from the fireball at an earlier stage of the fireball evolution at a higher temperature.

The application of the statistical approach outlined above to the productions of multistrange baryons  assumes implicitly that in every single collision all strange species can be produced irrespectively on how many kaons were produced in that particular collision, i.e. irrespectively on how many $s\bar{s}$ were produced. Hence, in this case  strangeness is conserved only on average.
However it must remain conserved in each collision event separately.
This implies that the combination of strange hadron species, which can be observed in the final state, depends on the number of $s$ quarks in the fireball. The chemical equilibrium conditions are thus
different for  events with different numbers of produced kaons.

To handle this complication it was suggested in~\cite{KTV12} to divide the totality of events with strangeness production into classes of events with one, two, three and so on, $s\bar{s}$ quark pairs created --- the so-called $n$-kaon events.  The probability of creation of exactly $n$ strange quark pairs is $P_{s\bar s}^{(n)}$. If the pairs are produced in initial nucleon-nucleon collisions their number follows the
probability distribution
$P_{s\bar s,{\rm in}}^{(n)}={w}_{\rm in}^n/(1+{w}_{\rm in})^{n+1}$, where
${w}_{\rm in}$ is the averaged multiplicity of the initial pair production. For $s\bar{s}$ pairs produced in hadrochemical reactions the probability is determined by the Poisson distribution
$P_{s\bar s, {\rm r}}^{(n)}  = w_{\rm r}^n \,e^{-w_{\rm r}} /n!$\,. When both mechanisms are operative we have $P_{s\bar s}^{(n)}=\sum_{i=0}^n C_n^i P_{s\bar s,{\rm in}}^{(i)}
P_{s\bar s, {\rm r}}^{(n-i)}$ with $C_n^i$ standing for the binomial coefficient.
The parameters $w_{\rm in}$ and $w_{\rm r}$ are constrained by the total multiplicity
of $K^+$ mesons observed in an inclusive collision $\mathcal{M}_{K^+}=\sum_{n} \langle n P_{s\bar{s}}^{(n)}\rangle/(1+\eta)$. The angular brackets stand for the averaging over collision impact parameter. The quantities $w_{\rm in}$ and $w_{\rm r}$ have different dependence on the system size and correspondingly on the impact parameter. The probability
$w_{\rm in}$ scales as $V_{\rm fo}^{\alpha}$ with $\alpha$ between 2/3 and 1,
depending on the penetration depth of incident nucleons.
Here $V_{\rm fo}$ is the fireball freeze-out volume determined by the freeze-out baryon density and the overlap volume of the initial colliding nuclei. On the other hand $w_{\rm r}$ grows with the volume of the system and the time the collision lasts. For ideal hydrodynamics, the only length scale is given by $V_{\rm fo}^{1/3}$ and it must determine also the lifetime~\cite{RI92} and hence  $w_{\rm r} \propto V_{\rm fo}^{4/3}$.

The statistical probability that strangeness will be released at freeze-out in a hadron of type $a$ with the mass $m_a$ is given by the standard Gibbs' formula
\begin{eqnarray}
P_a = z_S^{s_a}\,V_{\rm fo}\, p_a\equiv  z_S^{s_a} V_{\rm fo} \nu_a e^{B_a\frac{\mu_{B,{\rm fo}}}{T_{\rm fo}}}
\frac{m_a^2\, T_{\rm fo}}{2\pi^2} K_2\left(\frac{m_a}{T_{\rm fo}}\right )
,
\label{h-density}
\end{eqnarray}
where $B_a$ is the baryon number of the hadron, the degeneracy factor $\nu_a$ is determined by
the hadron's spin and isospin. The baryon chemical potential, $\mu_{B,{\rm fo}}$, at freeze-out is determined by the total baryon density~\cite{KT05,KTV12}.
The quantity $z_S$ in (\ref{h-density}) is a normalization factor. The factor $z_S^{s_a}$ follows from the requirement that the sum of production probabilities of different strange species and their combinations allowed in the finale state is equal to one. The factor $z_S^{s_a}$ depends on how many strange quarks are produced. Therefore, we introduce the notation $ P_a^{(n)} = (z_S^{(n)})^{ s_a}\,V_{\rm fo}\, p_a$\,, where the superscript $n$ indicates to which event class this probability and the $z_S$ factor belong:
In \emph{a single-kaon event} one $s$-quark can be released  as $\bar{K}$, $\Lambda$ or $\Sigma$. Hence, the normalization condition for the probabilities (\ref{h-density})
and the multiplicity $M_a^{(1)}$ of strange hadrons of type $a=\{\bar{K}, \Lambda, \Sigma\}$ are given by $z_S^{(1)}\,V_{\rm fo}\, (p_{\bar{K}}+p_{\Lambda}+p_{\Sigma})=1$ and
$M_{a}^{(1)}= g_a \, P_{s\bar{s}}^{(1)}\,P^{(1)}_a$, respectively.
The isospin factor $g_a$ takes into account the asymmetry in the yields of particles with various isospin projections induced by the global isospin asymmetry of the collision $\eta$.
In \emph{a double-kaon event} there can be one $\Xi$ baryon and all possible combinations of kaon and hyperon pairs. Therefore we have
$z_S^{(2)2}\,V^2_{\rm fo}\, (p_{\bar{K}}+p_{\Lambda}+p_{\Sigma})^2 + z_S^{(2)2}\,V_{\rm fo}\, p_{\Xi}=1$, and the multiplicities of singly-strange particles are
$M_{a}^{(2)} =g_{a}\, 2\,P_{s\bar{s}}^{(2)}\,P^{(2)}_{a}
(P^{(2)}_{\bar{K}}+P^{(2)}_{\Lambda}+P^{(2)}_{\Sigma})$, where for the $\Xi$ multiplicities we have $M_\Xi^{(2)} = g_\Xi\,P_{s\bar{s}}^{(2)}\,\,P^{(2)}_{\Xi}$\,.
The $\Omega^{-}$ baryons can be produced only in events with three and more kaons in the final state. Then the normalization condition is
$ z_S^{(3)3}\,V^3_{\rm fo}\,(p_{\bar{K}}+p_{\Lambda}+p_{\Sigma})^3
+ 3\,z_S^{(3)3} V^2_{\rm fo}(p_{\bar{K}}+p_{\Lambda}+p_{\Sigma})
p_\Xi +z_S^{(3)3}V_{\rm fo} p_{\Omega}=1$ and the multiplicities are
$ M_a^{(3)}=g_a\,3 P_{s\bar{s}}^{(3)} P^{(3)}_a\,
[\big(P^{(3)}_{\bar{K}}+P^{(3)}_{\Lambda}+P^{(3)}_{\Sigma}\big)^2 +  P^{(3)}_\Xi]
$ for the hadrons with one $s$ quark, for $\Xi$ baryon
$M_\Xi^{(3)}=g_\Xi\,3P_{s\bar{s}}^{(3)} \, P^{(3)}_\Xi\,(P^{(3)}_{\bar{K}}+P^{(3)}_{\Lambda}+P^{(3)}_{\Sigma})$
and for the  $\Omega$ baryon $M_\Omega^{(3)}=P_{s\bar{s}}^{(3)}\, P^{(3)}_\Omega$.
Having the normalization factors and the chemical potential at our disposal, we can calculate the multiplicity ratios of strange particles as functions of the freeze-out density and temperature.

\begin{figure}
\centering
\includegraphics[width=7cm]{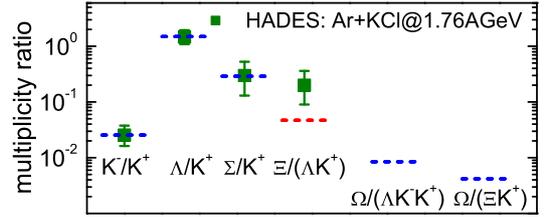}
\caption{
The strange particle ratios calculated in~\cite{KTV12} in comparison with the HADES data~\cite{HADES-Xi}.
}
\label{fig:HADES}
\end{figure}

This scheme was applied in~\cite{KTV12} to the description of multiplicity ratios of strange particles detected by the HADES collaboration in Ar+KCl collisions at $1.76\,A$GeV~\cite{HADES-Xi}. The main effect from the exact strangeness conservation
and volume averaging was
a reduction of the ratio $R_{\Xi/\Lambda/K^+}$ by factor $\sim 0.5$, the ratio $R_{\Omega/\Lambda/ K^{-}/K^+}$ by $\sim 0.2$ and the ratio $R_{\Omega/\Xi/K^+}$ by factor $\sim 0.4$ compared to the results obtained without exact strangeness conservation in each event. The inclusion of in-medium potentials for hyperons and $\bar{K}$ mesons, as discussed in Sect.~\ref{sec:medium} is important, leading to much better agreement with the experiment.
The resulting particle ratios are depicted in Fig.~\ref{fig:HADES} in comparison with the experimental ratios. We see that the calculations are in good agreement with the experiment for the ration of singly-strange particles. The multiplicity of $\Xi$ baryons is, however, strongly underestimated.

\begin{figure}
\centering
\includegraphics[width=7.9cm]{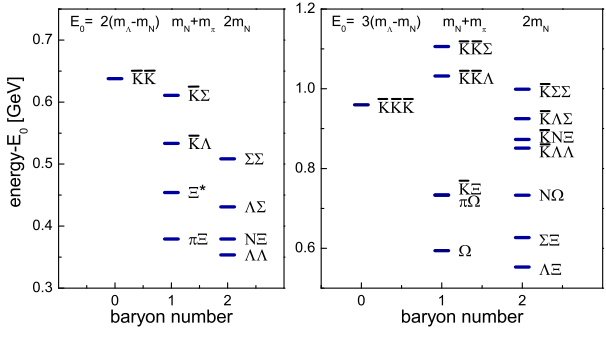}
\caption{
Mass spectrum of strange hadronic states with different baryonic numbers from the energy of non-strange ground states.}
\label{S-spec}
\end{figure}

Several possible sources of the $\Xi$ enhancement were discussed in~\cite{KTV12} with the conclusion that to get any substantial increase in the number of $\Xi$'s we have to assume that these baryons are not absorbed after being produced and their number is determined by the rate of direct production reactions, as, e.g., for dileptons.
There are
two types of reactions: the strangeness creation reactions, like $\bar{K}N\to K\Xi$, $\pi H\to K\Xi$, and strangeness recombination reactions, like $\bar{K}H\to \Xi\pi$ and $HH'\to \Xi N$. The former ones have very high thresholds and, therefore, are not operative at SIS energies.
Strangeness recombination reactions are secondary processes involving two strange particles. They can be promoted statistically. We argue that once several $s$ quarks appear in  the hadronic system it is more energetically
favorable to ``store'' them together in a multistrange state, $\Xi$ or $\Omega$. In Fig.~\ref{S-spec} we show the mass spectrum of strangeness --2 and --3 states in channels with different baryon numbers $B$. We see that in both single and double baryon channels the states with the $\Xi$ and $\Omega$ baryon are at the bottom of the spectrum. This means that $\Xi$s and $\Omega$s would play a role of a strangeness reservoir, being filled with a decrease of the temperature. The only sink is the reaction $\Xi N\to \Lambda\Lambda$ which has, however, a relatively small cross section~\cite{Polinder07}.\\[-7mm]

\section{Conclusions}

We reviewed a minimal statistical model in which the
total strangeness content of the fireball at freeze out is estimated using kaon multiplicity observed in HIC or calculated in hadro-chemical kinetic theory. We demonstrated how the strangeness conservation can be taken into account in each collision event separately. This effectively reduces the multiplicities of multistrange particles, $\Xi$ and $\Omega$. The model can be successfully applied for the description of singly-strange particle yields at SIS and AGS-SPS energies. In the former case the inclusion of in-medium effects improves the description sizeably. However relative $\Xi$ yields remain systematically underestimated for all energies. That may mean continuous production of
multistrange hyperons when they freeze out right after they are produced in the system.
For the best understanding of
the strangeness dynamics in the future experiments at NICA it would be wishful to measure with good precision not only kaons and antikaons but also several hyperon species including multistrange ones.\\[-7mm]

\begin{acknowledgement}
The work  was partially supported by grants APVV-0050-11, VEGA 1/0469/15 (Slovakia) and LG13031,  LG15001 (Czech Republic).
Computing was partially performed in the High Performance Computing Center of the Matej Bel University using the HPC infrastructure acquired in project ITMS 26230120002 and 26210120002 (Slovak infrastructure for high-performance computing) supported by the Research \& Development Operational Programme funded by the ERDF.
\end{acknowledgement}

\end{document}